\documentstyle[11pt]{article}
\begin{document}
{\setlength{\oddsidemargin}{1.4in}
\setlength{\evensidemargin}{1.4in} } \baselineskip 0.50cm
\begin{center}
{\LARGE {\bf Rotating dark matter solution \\ admitting dark fluid}}
\end{center}

\begin{center}
Ng. Ibohal \\
Department of Mathematics, Manipur University,\\
Imphal 795003, Manipur, INDIA.\\
E-mail: ngibohal@iucaa.ernet.in
\end{center}

\begin{center}
Ngangbam Ishwarchandra\ddag \quad and \quad K. Yugindro Singh\dag\\
Department of Physics, Manipur University,\\
Imphal - 795003, Manipur, India\\
E-mail: \ddag \,ngishwarchandra@gmail.com, \dag\,yugindro361@gmail.com
\end{center}
\date{}

\vspace*{.25in}
\begin{abstract}
In this paper we present an exact solution of Einstein's field equations, describing a rotating stationary axisymmetric dark matter, which is an extension of non-rotating dark matter solution of Paper I. We find that the rotating stationary solution is a non-asymptotic, Petrov type D space-time whose energy-momentum tensor admits a dark fluid having negative pressure and the energy equation of state parameter with minus sign.  We also find that the time-like vector fields of the matter distribution is expanding, shearing and rotating  with acceleration. It is also found that, due to the negative pressure, the energy-momentum tensor of the solution violates the strong energy condition leading to the repulsion of the gravitational field of the rotating space-time geometry. We also analyze the entropy and surface gravity for the horizon of the rotating dark matter solution. 
\end{abstract}

{\bf 1. Introduction}

\setcounter{equation}{0}
\renewcommand{\theequation}{1.\arabic{equation}}

In a recent paper [1], referred to as Paper I, an exact solution of Einstein's field equations has been derived, describing dark matter, whose energy-momentum tensor admits a non-perfect fluid having the negative pressure and the dark energy equation of state parameter $w=-1/2$.
The most striking property of this solution is that the metric of the solution describes both the background space-time structure and the dynamical aspects of the gravitational field in the form of the energy-momentum tensor. That is, the mass $m$ of the solution plays the role of both the curvature of the space-time ({\it non-flat}) as well as the source of the energy-momentum tensor with $T_{ab}\neq 0$ ({\it non-vacuum}) measuring the energy density and the negative pressure. This gives the information that when the mass $m$ of the dark matter solution is set to zero, the space-time will become the flat Minkowski space with vacuum structure $T_{ab}=0$. In the case of Schwarzschild solution, the mass plays only the role of curvature of the space-time geometry and cannot determine the energy-momentum tensor. Thus, the Schwarzschild solution is {\it non-flat vacuum} space-time with $T_{ab}=0$. Here the advantage of the dark matter solution over the Schwarzschild one is that it is {\it non-flat} and {\it non-vacuum} space-time metric.

Nowadays, various possible models have been suggested for the composition of dark matter in [2-11] and references there in. The observations of luminosity-redshift relation for the type Ia supernovae [12-14] suggest that the missing energy should possess negative pressure $p$ and the equation of state parameter $w=p/\rho$ [11]. It is emphasized the fact that the approach in the derivation of the dark matter solution [1] is merely based on the identification of the power $n=2$ of the Wang-Wu mass function [15] without any extra assumption. The identification of the power $n=2$ in the Wang-Wu mass function has first considered  in [1] and not been seen discussed before in the scenario of exact solutions of Einstein's field equations. This fact can be seen from
the mass function as a power series of the coordinate $r$ as
\begin{equation}
M(u,r)= \sum_{n=-\infty}^{+\infty} q_n(u)\,r^n,
\end{equation}
where $q_n(u)$ are arbitrary functions of $u$. Wang and Wu analyzed the mass function for generalization of {\sl non-rotating} Vaidya solution by choosing the function $q_n(u)$ corresponding to the number $n$ [15]. In [16] the mass functions has been utilized in rotating system and found the role of the number $n$ in generating rotating embedded solutions of Einstein's field equations. The roles of the power $n$ in the expansion series (1.1) for the known spherically axisymmetric solutions are cited as follows
[15, 16])
\begin{itemize}
  \item[(i)] $n = 0$ corresponds to the term containing mass of the vacuum Kerr family solutions such as Schwarzschild, Kerr.
  \item[(ii)] $n = -1$ purveys the charged term of Kerr family such as Reissner-Nordstrom, Kerr-Newman.
  \item[(iii)] $n = 1$ furnishes the global monopole solution [15].
  \item[(iv)] $n = 2$ provides the dark matter solution [1].
  \item[(v)] $n = 3$ contributes the de Sitter cosmological models, rotating and non-rotating [16].
\end{itemize}
These values of $n$ are used for studying stationary solutions (non-rotating and rotating). The non-stationary (time dependent) Vadya-Bonnor black holes, non-rotating [15] and rotating [16] can be obtained when $n = 0$ and $n =-1$ as Vaidya-Bonnor metric describes a charged solution. Non-stationary rotating as well as non-rotating de Sitter cosmological solutions can also
be obtained when $n=3$ [17]). In fact the Wang-Wu power series expansion of the mass functions happens to be the most convenient approach to generate new (embedded or non-embedded) solutions of Einstein's field equations, using the complex scalar fields developed in [16] for the mass function $M(u,r,\theta)$ in Newman-Penrose (NP) formalism [18]. The above identifications of power $n$ indicate that the case $n=2$ corresponds to the dark matter solution of Paper I, whose the line element is reproduced here for use as
\begin{eqnarray}
ds^2&=&(1-2mr)du^2+2du\,dr-r^2(d\theta^2+{\rm sin}^2\theta\,d\phi^2)
\end{eqnarray}
where the constant $m$ is considered to be the mass of the dark matter and is non-zero for the existence of the dark matter distribution. When $u$ = constant, the surface is the future directed null cone. This line-element will not be asymptotically flat when $r\rightarrow \infty$, and has a coordinate singularity at $r=(2m)^{-1}$ describing a Lorentzian horizon. The line element (1.2) is certainly different from Schwarzschild solution with $g_{uu}=1-2M/r$ having singularity at $r=2M$, $M$ being the Schwarzschild mass. The energy-momentum tensor for the above solution takes the following form
\begin{eqnarray}
T_{ab}=(\rho+p)(u_{a}u_{b}-v_{a}v_{b})-pg_{ab},
\end{eqnarray}
where $u_{a}$ is a time-like vector ($u_{a}u^{a}=1$) and $v_{a}$ is a space-like  ($v_{a}v^{a}=-1$). $g_{ab}$ is the metric tensor. The energy density $\rho$ and the pressure $p$ are found as
\begin{eqnarray}
&& \rho = {4\over Kr}m, \quad p = -{2\over Kr}m.
\end{eqnarray}
The parameter of the dark energy equation of state is the ratio of the pressure to the energy density and takes the constant value as
\begin{eqnarray}
\omega=\frac{p}{\rho}=-{1\over2}.
\end{eqnarray}
It is noted that the energy-momentum tensor (1.3) having negative pressure (1.4) does not describe a perfect fluid. Hence, we refer it to as dark fluid with the equation of state (1.5). The 4-velocity vector fields $u_a$ associated with (1.3) is expanding $\Theta \neq 0$, accelerating $\dot{u}_a\neq 0$ as well as shearing $\sigma_{ab} \neq 0$ with zero-twist. This means that the observer of the dark matter does not follow the path of the time-like geodesic as $u_{a;b}u^b\neq 0$. The non-geodesic condition is a striking feature of the solution to be interpreted as an accelerated expansion of space-time geometry (1.2). The energy-momentum tensor for the dark fluid violates the strong energy condition $p \geq 0$, $\rho + p \geq 0$. The violation of strong energy condition is due to the negative pressure of the matter field content in the space-time geometry, and is not an assumption to obtain the solution (like other dark matter models mentioned in Sahni [2]). This violation indicates that the gravitational field of the solution is repulsive leading to the accelerated expansion of the space-time geometry [1]. The expansion of the dark matter solution with acceleration is consistent with the observational data [12-14]. Here it is to mention the fact that the energy-momentum tensor for the electromagnetic field in Ressner-Nordstrom and Kerr-Newman solutions does not violate the strong energy condition $p \geq 0$, $\rho + p \geq 0$, since $p$ takes positive value [16] causing attractive force of gravitational field, rather then repulsive as in the case of dark matter [1].

From the analysis of the dark matter solution in Paper I, we find that the dark energy density is only ascribed from the mass of the dark matter. It is the fact that without the mass of the dark matter one cannot determine the energy density and the negative pressure in the energy-momentum tensor distribution (1.3) in order to obtain the dark energy equation of state $w = -1/2$. This means that the negative pressure and the energy density of the dark fluid associated with the energy-momentum tensor are measured by the mass that produces the gravitational field in the space-time geometry of the dark matter. Hence, we may conclude that the solution (1.2) can explain the essential part of Mach's principle -- ``The matter distribution influences the space-time geometry'' [19]. Here it is also emphasized that usually dark matter and dark energy are considered to be separate components of the Universe [2-11]. Arbey [20] has unified both components into a single ``dark fluid'' in order to explain the cosmological observation, based on a complex scalar field. In this regard the solution of Paper I indicates that dark matter and dark energy  cannot be regarded as separate components, since the {\it dark energy} with the negative pressure is produced by the mass of {\it dark matter}, {\it i.e.} the same mass $m$ plays the important key role for the existence of both.

The space-time of the dark matter has an event horizon at $r=(2m)^{-1}$. Eventually, the area, entropy as well as surface gravity associated with the horizon have been analyzed. It is shown in [1] that the surface gravity is directly proportional to the mass of the dark matter. This indicates that the existence of the mass implies the existence of its surface gravity and the temperature of the dark matter on the horizon. The existence of dark matter horizon is in consistency with the cosmological horizon of de Sitter space with constant $\Lambda$ [21], which is considered to be a common example of dark energy with the parameter $w= - 1$ [3-11]. This parameter of equation of state is also satisfied for both the cosmological constant $\Lambda$ as well as the cosmological function $\Lambda(u)$ of the {\it rotating} and {\it non-rotating} de Sitter solutions [17]. It is quite interesting to specify that the approximate size of the dark matter mass is found, in the length scale $r>10^{60}$ suggested by Bousso [5], as $m<(1/2)\times 10^{-60}$, which is doubly bigger than the size of the cosmological constant $|\Lambda|\leq3\times10^{-120}$ with the horizon $r_\Lambda=\sqrt{3/\Lambda}$. This incredibly small size of the mass may be one reason why dark matter is not been able to observe, however its gravitational effect is being recognized in the form of dark energy in the dark halo of the Universe. Another possible reason may be that it behaves like a invisible black hole with event horizon. We believe that the non-rotating solution [1] may provide an example of an {\it invisible dark matter} which can produce dark energy with the equation of state parameter $w=-1/2$ in the accelerated expanding space-time geometry. However, the non-rotating dark matter solution of Paper I cannot explain the gravitational field of a rotating system. Therefore, this shows the requirement for developing a rotating dark matter solution of Einstein field equations in order to understand the complete nature of such invisible dark matter in the Universe.

Thus, the purpose of the paper is to generalize the rotating dark matter  solution from the non-rotating one of Paper I, and then to investigate the nature of the rotating system. For instance, the rotating vacuum Kerr solution is an extension of the non-rotating Schwarzschild vacuum solution, and the Kerr-Newman solution is the charged rotating black hole which is an extension of the non-rotating Reissner-Nordstrom black hole. It is the fact that all astrophysical objects in the Universe are rotating on their own axes with very distinct physical properties. Hence, such a generalized rotating solution may provide more information about the gravitational field for better understanding the nature of dark matter in the Universe.

The paper is organized as follows: In Section 2 the derivation of rotating dark matter solution of Einstein's equations of gravitational field has been shown. We find that the mass of the rotating dark matter solution proposed here describes the gravitational field of the space-time geometry and also determines the matter distribution with negative pressure, whose dark energy equation of state has the negative value $w=-1/2$ at the poles $\theta=\pi/2$ and $\theta=3\pi/2$; otherwise it takes the functional form with minus sign  as in (2.13) below. This is one of the curious properties of the rotating solution that the non-rotating one (1.2) cannot explain this situation. For the non-rotating dark matter $a=0$ it is always with the constant value $w=-1/2$ as seen from the above. The variation of the value of $w$ is consistent with those of the cosmological constant $\Lambda$ [16] and function $\Lambda(u)$ [17] of rotating de Sitter spaces. So we refer the solution obtained here to as {\it rotating dark matter} solution. The energy produced by this dark matter with the equation of state parameter is also regarded as {\it rotating dark energy}. However, the energy-momentum tensor of the dark matter distribution with negative pressure does not describe a perfect fluid, which can be seen in Section 2. This non-perfect fluid distribution is in accord with the remark of Islam [22] -- ``it is not necessarily true that the field is that of a star made of perfect fluid''. We also find that the dark matter solution has an event horizon. Consequently, we present the area, entropy and surface gravity for the horizon. The existence of the dark matter horizon discussed here is in accord with the cosmological horizon [21] of the rotating de Sitter spaces with constant $\Lambda$ having the equation of state parameter $w=-1$ at the poles. We conclude the paper in Section 3 with considerable remarks and evolution of the rotating solution. Thus, we summarize the results of the paper in the following theorem:
\newtheorem{theorem}{Theorem}
\begin{theorem}
The rotating stationary dark matter solution is a non-asymptotic, non-vacuum, Petrov type D space-time, whose stress energy-momentum tensor possesses a dark fluid having a negative pressure and dark energy equation of state parameter with minus sign.
\end{theorem}
\begin{theorem}
The 4-velocity vector of the rotating dark fluid having negative pressure is non-geodesic, expanding, accelerating, shearing as well as rotating.
\end{theorem}
Theorem 1 shows the physical interpretation of the solution that the space-time metric for the solution is not asymptotically flat when $r\rightarrow \infty$, and is Petrov type D with the non-zero component $\psi_2 \equiv -C_{abcd}\,\ell^a\,m^b\,\overline{m}^c\,n^d \neq 0$ of Weyl tensor $C_{abcd}$. The energy-momentum tensor associated with the solution has a negative pressure with the dark energy equation of state parameter $w(r,\theta)$ in (2.13). It is to emphasize that the Theorem 1 above is regarded as a generalization of the theorem of Paper I -- ``{\it The non-rotating stationary dark matter solution is a non-asymptotic,
non-vacuum, conformally flat space-time, possessing a stress energy-momentum tensor of a dark fluid with negative pressure and dark energy equation of state parameter $w=-1/2$}''. Theorem 2 indicates the kinematic properties of the dark fluid content in the rotating axisymmetric space-time geometry of the dark matter. The presentation of this article is precisely based on mathematical
calculations for obtaining a new {\it rotating stationary} (time independent) exact solutions of Einstein's field equations without any extra assumption. We do not attempt to use any observational data available at present. The Newman-Penrose (NP) formalism [18] in $(-2)$ signature is adopted as mathematical tools in this paper.

\vspace*{.15in}
{\bf 2. Rotating dark matter solution}
\setcounter{equation}{0}
\renewcommand{\theequation}{2.\arabic{equation}}
\vspace*{.15in}

In this section we extend the {\it non-rotating} solution of Paper I to a {\it rotating} dark matter solution, as the rotating space-time metric has considerable potential application in astrophysical studies -- all objects in astrophysics are rotating on their own axes. For instance, the non-rotating Schwarzschild vacuum black hole is generalized in the form of rotating Kerr vacuum one which has great application in the study of black holes with rotating configuration. Hence it is the main aim to derive a generalized version of the non-rotating solution of [1] in the form of rotating stationary (time independent) axisymmetric dark matter solution for better understanding the nature of dark matter problem. For this purpose we choose the Wang-Wu function $q_{n}(u)$ in $(1.1)$ as in Paper I
\begin{eqnarray}
\begin{array}{cc}
q_n(u)=&\left\{\begin{array}{ll}
m, &{\rm when}\;\;n=2\\
0, &{\rm when }\;\;n\neq 2,
\end{array}\right.
\end{array}
\end{eqnarray}
such that the mass function (1.1) takes the form
\begin{equation}
M(u,r)\equiv \sum_{n=-\infty}^{+\infty} q_n(u)\,r^n =m r^2,
\end{equation}
where $m$ is constant and $u=t-r$ is the retarded time coordinate.
Then using the mass function (2.2) in general rotating metric presented
in equation (6.4) of [16], we obtain a rotating metric,
describing a {\sl stationary} solution of dark matter in the null
coordinates system $(u,r,\theta,\phi)$. The line element is found as
\begin{eqnarray}
ds^2&=&\{1-2r^3mR^{-2}\}du^2 +2du\,dr
+4a{r^3mR^{-2}}\,{\rm sin}^2\theta\,du\,d\phi \cr &&
-2a\,{\rm sin}^2\theta\,dr\,d\phi -R^2d\theta^2-\{(r^2+a^2)^2
-\Delta a^2{\rm sin}^2\theta\}R^{-2}{\rm
sin}^2\theta\,d\phi^2,
\end{eqnarray}
where $R^2=r^2+a^2{\rm cos}^2\theta$ and $\Delta =r^2-{2r^3\,m}+a^2$.
Here the constant $a$ describes the rotational parameter. When $a$ is set to zero, then the line element (2.3) will reduce to non-rotating dark matter solution of [1]. This space-time metric is invariant to the simultaneous inversion of the retarded time $u \rightarrow -u$ and the azimuthal angle $\phi\rightarrow -\phi$ and also the metric coefficients be independent of $u$ and $\phi$. These are the requirement characters for a metric to be {\it stationary} and {\it axisymmetric}. The line element is non-asymptotic when $r\rightarrow \infty$. However, it has a singularity when $\Delta=0$ having three roots given in (2.28) below. The line element is certainly different from the Kerr metric with $\Delta^*=r^2-2rM+a^2$, where $M$ is the mass of the Kerr black hole having two roots for $\Delta^*=0$.

Then the covariant complex null tetrad vectors for the metric can  be chosen as follows
\begin{eqnarray}
&&\ell_a=\delta^1_a -a\,{\rm sin}^2\theta\,\delta^4_a,\cr
&&n_a=\frac{\Delta}{2\,R^2}\,\delta^1_a+ \delta^2_a
-\frac{\Delta}{2\,R^2}\,\,a\,{\rm sin}^2\theta\,\delta^4_a,\\
&&m_a={1\over\surd 2R}\,\{ia\,{\rm
sin}\,\theta\,\delta^1_a-R^2\,\delta^3_a -i(r^2+a^2)\,{\rm
sin}\,\theta\,\delta^4_a\}.
\nonumber
\end{eqnarray}
where $R=r+ia\cos\theta$. Here $\ell_a$,\, $n_a$
are real null vectors and $m_a$ is complex with the normalization
conditions $\ell_an^a= 1 = -m_a\bar{m}^a$ and other inner products are vanished. These null tetrad vectors will be utilized to calculate Newman-Penrose quantities including spin coefficients, tetrad components of Ricci and Weyl tensors. From the tetrad components of Ricci tensor
we calculate the energy-momentum tensor by virtue of Einstein's field equations. The Newman-Penrose spin coefficients [18], which are twelve complex quantities,
representing the 24 Christoffel symbols involved in calculating
the Ricci tensor are given below
\begin{eqnarray}
&&\kappa^* =\sigma =\lambda =\epsilon =\nu=0,\cr
&&\rho^* = -{1\over \bar{R}}, \quad  \mu^* = -{\Delta\over
2\bar{R}R^{2}}, \cr &&\alpha = {2ia-R\cos\theta\over2\surd2 \bar{R}\bar{R}\sin\theta},\quad  \beta ={1\over2\surd2 R}\cot\theta,\cr &&\pi={i\,a\,{\rm
sin}\,\theta\over{\surd 2\bar{R} \,\bar{R}}},\;\;\;
\tau=-{i\,a\,{\rm sin}\,\theta\over{\surd 2R^2}}, \\
&&\gamma = {1\over 2\bar{R}R^{2}}\{(r-3mr^{2})\bar{R}-\Delta\}.\nonumber
\end{eqnarray}
The Ricci scalars of the metric are obtained as follows:
\begin{eqnarray}
 &&\phi_{11}\equiv-\frac{1}{4}R_{ab}(l^{a}n^b+m^a\bar{m}^b)=\frac{mr}
{2R^2R^2}(r^2-3a^2\rm{cos}^2\theta)\cr
 &&\Lambda^*\equiv \frac{1}{24}R_{ab}g^{ab} =\frac{1}{2R^2}rm
\end{eqnarray}
where $R_{ab}$ is the Ricci tensor.
The complex tetrad component of the Weyl tensor $C_{abcd}$ determining gravitational field of the space-time (2.3) is found as
\begin{eqnarray}
&&\psi_2\equiv -C_{abcd}l^am^b\bar{m}^cn^d=-{m\over\bar{R}\bar{R}\bar{R}}iar\,\cos\theta.
\end{eqnarray}
This expression of Weyl curvature scalar $\psi_2$ shows the coupling of the rotational parameter $a$ with the mass $m$ of the dark matter. When the rotational parameter takes the limit $a=0$, the Weyl scalar $\psi_2$ vanishes to lead to the conformally flat space-time as in non-rotating dark matter solution of Paper I. From (2.7) we find that the rotating stationary dark matter (2.3) is type D in the Petrov classification of space-time with a repeated principal null congruence $\ell_a$ (2.4), which is geodesic $(\kappa=\epsilon=0)$, shear free $(\sigma=0)$, expanding ($\hat{\theta}\equiv
{1\over2}\ell^a_{\,;a}$) as well as non-zero twist
($\hat{\omega}^2 \equiv \frac{1}{2}\ell_{[a;\,b]}\ell^{a;\,b}$), where
\begin{eqnarray}
&&\hat{\theta}\equiv -{1\over2}(\rho^* + \bar{\rho}^*) ={r\over R^2}, \cr
&&\hat{\omega}^{2}\equiv-{1\over4}(\rho^* - \bar{\rho}^*)^2
={{a^2\cos^2\theta}\over {R^2\,R^2}}.
\end{eqnarray}

We find the energy-momentum tensor describing matter field for the above rotating stationary space-time (2.3) as follows:
\begin{eqnarray}
T_{ab} &=& 2\,\rho\,\ell_{(a}\,n_{b)}
+2\,p\,m_{(a}\bar{m}_{b)},
\end{eqnarray}
where the density and pressure are found as
\begin{eqnarray}
&&\rho = {4r^3m\over
KR^2R^2}, \quad p = -{2rm\over KR^2R^2}\{r^2+3a^2{\rm
cos}^2\theta\}.
\end{eqnarray}
Because of the involvement of the rotational parameter $a$ in $R^2=r^2+a^2cos^2\theta$, we find the difference in the values of $\rho$ and $p$ in (2.10) from those given in (1.4) of non-rotating solution. The energy-momentum tensor (2.9) can be expressed in the form of (1.3) using the orthonormal tetrad vectors $\{u_a,v_a,w_a,z_a\}$ given in Paper I. From (2.10) it is found that the stress energy tensor (2.9) admits the {\it weak energy condition} $(\rho>0$, $\rho+p>0)$ as well as {\it dominant energy condition} $(\rho^2>0$, $\rho^2-p^2>0)$. However, the {\it strong energy condition} $(p>0$, $\rho+p>0)$ is violated as the pressure $p$ has negative value in (2.10). This violation may lead to a non-attractive (repulsive) gravitational force of a rotating dark matter solution (2.3) as pointed in [23].

The energy-momentum tensor of the solution (2.3) satisfies the energy conservation equations [17] formulated in Newman-Penrose formalism [18].
\begin{eqnarray}
T^{ab}_{\;\;\,;b}=0.
\end{eqnarray}
This establishes the fact that the space-time metric (2.3) describing dark matter is an exact rotating stationary solution of Einstein's field equations. The trace of energy momentum tensor $T_{ab}$ (2.9) is found as
\begin{equation}
T=2(\rho-p)={12r\over KR^2}m.
\end{equation}
Here it is observed that $\rho - p>0$ for rotating
stationary of dark matter model. The energy-momentum
tensor (2.9) does not describe a perfect fluid, i.e. for a
non-rotating perfect fluid $T^{(\rm {pf})}_{ab} =
(\rho+p)u_a\,u_b-p\,g_{ab}$ with unit time-like vector $u_a$ and
trace $T^{(\rm {pf})}=\rho-3p$, which is different from the one
given in (2.12).

\vspace*{.15in}
{\bf The equation of state:}
The equation of the state for the dark energy is the ratio of the pressure to the energy density, and is found for the rotating solution (2.3) with (2.10) as
\begin{eqnarray}
w={p\over\rho}=-{1\over2}\Big\{{1\over r^2}(r^2+3a^2\cos^2\theta)\Big\},
\end{eqnarray}
which shows the advantage of the rotating solution over the non-rotating one (1.2).
Here we have seen the difference between the non-rotating (1.2) as well as the rotating dark matter solution (2.3) that the non-rotating one has the constant parameter $w=-1/2$ of the energy equation of state; but that of the rotating solution takes the functional form (2.13), which may attend the constant value $w=-1/2$, if
\begin{equation}
{1\over r^2}(r^2+3a^2\cos^2\theta)=1.
\end{equation}
This is true at the poles $\theta=\pi/2$ and $\theta=3\pi/2$ with $a\neq 0$, showing that the solution (2.3) describes the rotating dark matter producing dark energy with the equation of state parameter (2.13). This indicates the necessity for the analysis of a rotating dark matter providing more information than the non-rotating one [1]. It is noted that the equation (2.13) is always satisfied for the non-rotating dark matter solution of [1] with $a=0$. However, the fact that $w$ is not always equal to $-1/2$ for every value of $\theta$ in (2.13), except $\theta=\pi/2$ and $3\pi/2$, suggests the variation of equation of state $w$ depending on the rotational configuration of the solution with $a\neq 0$. It is the fact that we need not necessarily concentrate with {\it non-rotating solution} as in [1], and now we have a rotating solution which may provide the complete structure of dark matter in the Universe. The functional parameter $w(r,\theta)$ (2.13) explains the main development of the rotating dark matter (2.3) producing dark energy, rather then the constant parameter $w$ (1.5) for non-rotating solution. From (2.7), (2.9) and (2.13), we come to the conclusion of the proof of Theorem 1, stated in the introduction above.

\vspace*{.15in}
{\bf Kerr-Schild ansatz} :
Let us express the line-element (2.3) of the dark matter space-time in Kerr-Schild ansatz in order to show that it is certainly a solution of Einstein's field equations. For this purpose, we write the line element (2.3) as
\begin{eqnarray}
ds^2&=&dt^2-dr^2-R^2d{\theta}^2-(r^2+a^2)\sin^2\theta\,d\phi^2
-2a\sin^2\theta\,dr\,d\phi \cr && -\frac{2mr^3}{R^2}\{du-a\sin^2\theta\,d\phi\}^2
\end{eqnarray}
where $dt=du+dr$.
This is of the Kerr-Schild ansatz on the Lorentzian flat background
\begin{eqnarray}
g_{ab}^{\rm dm}=g_{ab}^{\rm L}-{2mr^3\over R^2}\ell_a\ell_b,
\end{eqnarray}
with the null vector $\ell_a$ given in (2.4) and the Lorentzian flat metric
\begin{eqnarray}
g^{\rm L}_{ab}=\delta^t_a\delta^t_b-\delta^r_a\delta^r_b-R^2\delta^\theta_a \delta^\theta_b-(r^2+a^2)\sin^2\theta\delta^\phi_a\delta^\phi_b
-2a\sin^2\theta\delta^r_a\delta^\phi_b.
\end{eqnarray}
Now by the following transformation
\begin{eqnarray}
dt=du+\frac{r^2+a^2}{\Delta}dr, \quad
d\phi=d\tilde{\phi}+\frac{a}{\Delta}dr,
\end{eqnarray}
we express the line-element (2.15) in Boyer-Lindquist coordinate system for future use as
\begin{eqnarray}
ds^2&=&\frac{\Delta}{R^2}\{dt-a \sin^2\theta\,d\phi\}^2 -\frac{\sin^2\theta}{R^2}\{(r^2+a^2)d\phi-a dt\}^2 \cr
&&-\frac{R^2}{\Delta}dr^2-R^2d\theta^2,
\end{eqnarray}
where `$\sim$' is being removed from $d\tilde{\phi}$ for simplicity.
The line-element (2.19) can also be expressed in a very familiar form
\begin{eqnarray}
ds^2&=&\frac{1}{R^2}\{\Delta-a^2\sin^2\theta \}dt^2+\frac{4am}{R^2}r^3\sin^2\theta\, dt\,d\phi \cr
&&-\frac{R^2}{\Delta}dr^2-R^2d\theta^2-\frac{1}{R^2}\{(r^2+a^2)^2-{\Delta} a^2\sin^2\theta\}\sin^2\theta\,d\phi^2.
\end{eqnarray}
where $\Delta=r^2-2mr^3+a^2$.
This form of the space-time metric is quite useful for working in the Boyer-Lindquist coordinate system $(t,r,\theta,\phi)$. For example, we observe that the line element has a singularity when
\begin{equation}
\Delta \equiv r^2-{2mr^3}+a^2=0.
\end{equation}
This situation cannot be seen directly in (2.3). We can also find the {\it stationary limit} $g_{tt}=0$, which implies that
\begin{equation}
r^2-{2mr^3}+a^2\cos\theta=0.
\end{equation}
Here we have seen that the singularity $\Delta=0$ and the stationary limit $g_{tt}=0$ coincide at the poles $\theta=\pi/2$ and $\theta=3\pi/2$. One important point is to mention that the functional parameter $w(r,\theta)$ (2.13) of the energy equation of state takes the constant value $w=-1/2$ at these poles. This situation cannot be observed in the case of non-rotating dark matter solution (1.2) of the Paper I.

In order to express the rotating solution in Kerr-Schild anzatz based in Minkowski flat background, let us use the following transformation of coordinates [24,25].
\begin{eqnarray}
&&x=(r\,{\rm
cos}\phi + a\,{\rm sin}\phi)\,{\rm sin}\theta, \quad  y=(r\,{\rm
sin}\phi - a\,{\rm cos}\phi)\,{\rm sin}\theta, \cr &&z=r\,{\rm
cos}\theta,\;\;\; t=u+r,
\end{eqnarray}
where $r$ is defined in terms of $x, y,  z$ as
\begin{eqnarray}
&&r^4-(x^2+y^2+z^2-a^2)r^2-a^2z^2=0.
\end{eqnarray}
The metric (2.3) can be expressed in the coordinate system
 $(t, x, y, z)$ as
\begin{eqnarray}
ds^2=dt^2-dx^2-dy^2-dz^2-\frac{2mr^5}{r^4+a^2z^2}(\ell_{a}dx^a)^2,
\end{eqnarray}
where the null vector $\ell_{a}$ (2.4) is being used to obtain
\begin{eqnarray*}
\ell_{a}dx^a=dt-
{1\over(r^2+a^2)}\,\{r(xdx+ydy)+a(xdy-ydx)\}-{1\over r}\,zdz.
\end{eqnarray*}
The above transformed metric (2.25) is of the Kerr-Schild form
\begin{eqnarray}
&&g_{ab}^{\rm dm}=\eta_{ab}+2H(x,y,z)\ell_{a}\ell_{b},
\end{eqnarray}
where $\eta_{ab}$ is the Minkowski flat metric, $g_{ab}^{\rm dm}$ denotes the metric tensor for the rotating dark matter solution, and
\begin{eqnarray}
H(x,y,z)=-\frac{r^5m}{r^4+a^2z^2}.
\end{eqnarray}
$\ell^a$ is null with respect to both $\eta_{ab}$ and $g^{\rm dm}_{ab}$.
The line element (2.25) is regular at every point except at $r=0$, and $z=0$. The two Kerr-Schild forms in different coordinate systems (2.16) and (2.26) establish the conclusion that the rotating stationary axisymmetric solution (2.3) is certainly a solution of Einstein's equations of gravitational field. By virtue of these Kerr-Schild forms, we may refer the rotating dark matter (2.3) to as {\it dark matter solution is embedded into the flat background} or simply {\it dark matter in flat background}.

\vspace*{.15in}
{\bf Surface gravity}: Here let us show that the rotating dark matter metric has an event horizon at the point $r=r_1$, which is a real root of equation (2.21) describing singularity of the solution. In fact the equation (2.21) has three roots $r_{1}$, $r_{2}$ and $r_{3}$; $r_1$ is real and $r_2$ and $r_3$ are complex conjugate of each other and are found as
\begin{eqnarray}
&&r_1=\frac{1}{6m}\{1-h-h^{-1}\}, \\
&&r_2=\frac{1}{6m}\Big\{1+\frac{1}{2h}(1+i\sqrt{3})+\frac{1}{2}(1-i\sqrt{3})h\Big\},\quad {\rm and} \quad r_3=\bar{r}_2, \nonumber
\end{eqnarray}
where
\begin{equation}
h^3=\{-1-54a^2m^2+6am\sqrt{3(1+27a^2m^2)}\,\}
\end{equation}
These roots have the following relation
\begin{eqnarray*}
(r-r_1)(r-r_2)(r-r_3)= -\,\frac{1}{2m}\{r^2-2mr^3+a^2\}.
\end{eqnarray*}
We are interested only in the real root $r=r_1$ as the complex roots $r_2$, $r_3$ have unphysical meaning. When $a=0$, we have $h=-1$ and the real root becomes $r_1=(2m)^{-1}$, showing the consistency with that of the non-rotating solution (1.2). The area of the dark matter at event horizon $r=r_1$ is found as
\begin{eqnarray}
{\cal A} &=&
\int_{0}^{\pi}\int_{0}^{2\pi}\sqrt{g_{\theta\theta}g_{\phi\phi}}
\,d\theta\,d\phi\,\Big|_{r=r_1} = 4\pi\{r_{1}^2 +a^2\}.
\end{eqnarray}
According to Bekenstein-Hawking area-entropy formula ${\cal
S}={\cal A}/4$ [26], we find the entropy associated with $r=r_1$ as
\begin{eqnarray}
{\cal S}=\pi\{r_{1}^2 +a^2\}.
\end{eqnarray}
The gravity of the dark matter horizon is determined by the surface gravity which is defined as $\kappa n^a= n^b\nabla_{b}n^a$ in [24]. Here the null vector $n^a$ given in (2.4) above is parameterized by the coordinate $u$, such that $d/du=n^a\nabla_{a}$, and has the normalization condition $\ell_an^a=1$ with the null vector $\ell_a$. The surface gravity
$\kappa$ of the horizon associated with $r=r_{1}$ is found as
\begin{eqnarray}
\kappa=-(\gamma+\bar\gamma)=-\frac{r}{R^2}(1-3mr)\Big|_{r=r_1}.
\end{eqnarray}
where $R^2=r^2+a^2\cos^2\theta$ and $\gamma$ is the spin coefficient given in (2.5). Here we do not intend to substitute the value of $r_1$ in (2.30-2.32) as these equations do not have simplified forms. Instead, we look at the following case of extreme dark matter.

\vspace*{.15in}
{\bf Extreme dark matter}: Like extreme black holes [26], we shall discuss a case of extreme dark matter when $1+27a^2m^2=0$ in (2.29). In this case we have the quantity $h=1$, which leads to the real root $r_1=-(6m)^{-1}$. Eventually, we find the area and entropy for the extreme dark matter as follows
\begin{eqnarray}
{\cal A}=\pi a^2, \quad {\rm and} \quad {\cal S} = \frac{\pi}{4}a^2,
\end{eqnarray}
where $a$ is related with the mass $m$ by virtue of the equation $1+27a^2m^2=0$. Then the surface gravity for this case is found as
\begin{eqnarray}
\kappa=27m(3-4\cos^2\theta)^{-1},
\end{eqnarray}
measuring directly from the mass $m$ of the extreme dark matter, but can never be vanished for the existence of the dark matter. The functional parameter (2.13) of the energy equation of state becomes
\begin{equation}
w=-\frac{1}{2}\{1-4\cos^2\theta\}
\end{equation}
at $r_1=-(1/6m)$ of the extreme dark matter. This still retains the characteristic feature of the rotating dark matter having constant parameter $w=-1/2$ at the poles $\theta=\pi/2$ and $\theta=3\pi/2$.

\vspace*{.15in}
{\bf Generalized Chandrasekhar's Theorem}: Let us show the fact that the rotating dark matter solution (2.3), which is non-vacuum $(T_{ab} \neq 0)$ Petrov type $D$ space-time, satisfies the condition of generalized Chandrasekhar's theorem [27]. Chandrasekhar [25] has established a relation of spin coefficients $\rho^*$, $\mu^*$, $\tau$, $\pi$ in the case of an affinely parameterized geodesic vector, generating an
integral which is constant along the geodesic in a {\sl vacuum} $(T_{ab}= 0)$
Petrov type $D$ space-time
\begin{equation}
{\rho^*\over \bar{\rho}^*}={\mu^*\over\bar{\mu}^*}={\tau\over\bar{\pi}}
={\pi\over\bar{\tau}}.
\end{equation}
The original derivation of this relation is necessarily based on the vacuum Petrov
type $D$ space-time with Weyl scalars $\psi_2\neq 0$, $\psi_0=\psi_1=\psi_3=\psi_4=0$ and
vanishing of the tetrad components of Ricci tensor $\phi_{01}=\phi_{02}=\phi_{10}
=\phi_{20}=\phi_{12}=\phi_{21}=\phi_{00}=\phi_{22}=\phi_{11}=\Lambda=0$.
In [27] a complex tetrad component $\chi_1={1\over 2}
f_{ab}(\ell^a\,n^b+\overline{m}^a\,m^b)$ of Killing-Yano tensor $f_{ab}$ has been introduced in the above relation (2.36) as
\begin{equation}
{\rho^*\over \bar{\rho}^*}={\mu^*\over\bar{\mu}^*}={\tau\over\bar{\pi}}
={\pi\over\bar{\tau}}=-{\bar{\chi}_1\over\chi_1},
\end{equation}
where $f_{ab}$ is the skew-symmetric
tensor $f_{[ab]}$ satisfying the   Killing-Yano equations
\begin{equation}
f_{ab;c}+f_{ac;b}=0.
\end{equation}
The importance of KY tensor in
General Relativity seems to lie on Carter's remarkable
result [24] that the separation constant of Hamilton-Jacobi
equation (for charged orbits) in the Kerr space-time gives a
fourth constant. In fact, this constant arises from the scalar
field $K_{ab}v^a\,v^b$ which has vanishing divergent along a unit
vector $v^a$ tangent to an orbit of charged particle. Here $K_{ab}
=f_{ma}f^m_b$ is a symmetric Killing tensor. The KY tensor is skew-symmetric of rank two and has six real components. These six components can be expressed by three complex tetrad components $\chi_0$, $\chi_1$, $\chi_2$. Due to the vanishing of Weyl scalars $\psi_0=\psi_1=\psi_3=\psi_4=0$ for a Petrov type D space-time $\psi_2\neq 0$, the tetrad components $\chi_0$, $\chi_2$ of Killing-Yano tensor become zero. Then the remaining tetrad component $\chi_1$
can be solved from the Killing-Yano equations (2.38) transcribed in NP formalism in [27], and is found as
\begin{equation}
\chi_1=iC(r-i\,a\,\cos\theta)
\end{equation}
where $C$ is a real constant. Using this in (2.37) we obtain the relation as
\begin{equation}
{\rho^*\over \bar{\rho}^*}={\mu^*\over\bar{\mu}^*}={\tau\over\bar{\pi}}
={\pi\over\bar{\tau}}=-{\bar{\chi}_1\over\chi_1}
={R\over\bar{R}},
\end{equation}
where $R=r+i\,a\,\cos\theta$ and the spin coefficients $\rho^*$, $\mu^*$,
$\tau$, $\pi$ are given in (2.5) for the dark matter solution. Then
the Killing-Yano tensor  for the solution is found
\begin{equation}
f_{ab}=4a\cos\theta\,n_{[a}\ell_{b]}+4ir\,m_{[a}\bar{m}_{b]},
\end{equation}
and  accordingly, the Killing tensor $K_{ab}=f_{ma}\,f^m_b$ becomes
\begin{equation}
K_{ab}=-8\{a^2\cos^2\theta\,\ell_{(a}n_{b)}+r^2m_{(a}\bar
{m}_{b)}\}.
\end{equation}
We observe that the non-vacuum rotating dark matter solution is one of the examples which satisfy the generalized Chandrasekhar's relation (2.37).
The non-vacuum rotating stationary solutions, like Kerr-Newman, Kerr-Newman-de Sitter, negative mass naked singularity embedded into de Sitter space [28] and the dark matter solution discussed here are all Petrov type $D$ space-times and admit the relation (2.37). Hence, we conclude, without any lost of generality, that the generalized relation (2.37) with Killing-Yano scalar $\chi_1$ holds true for {\it every rotating stationary Petrov type $D$ solution, vacuum or non-vacuum}, admitting the KY (2.41) and Killing (2.42) tensors.

\vspace*{.15in}
{\bf Properties of time-like vector}: Here we shall discuss the nature of the 4-velocity vector $u^a$, which is time-like, appeared in the energy-momentum tensor for the dark fluid, if expressed in terms of orthonarmal tetrad vector fields $\{u_a,v_a,w_a,z_a\}$ as in (1.3), for the rotating dark matter solution (2.3) with (2.10). In fact the 4-velocity vector fields $u^a$ explain the kinemetic properties of a fluid distribution whether the fluid flow is expanding ($\Theta=u^a_{\:\,;a} \neq 0$), accelerating ($\dot{u}_a=u_{a;b}u^b\neq 0$), shearing $\sigma_{ab}\neq 0$ or rotating ($w_{ab}\neq 0$). For example, the dark matter of Paper I is expanding, shearing, accelerating but non-rotating. The key role of the rotational parameter $a$ of the stationary solution (2.3) can be seen in the following quantities. The expansion scalar $\Theta$ and acceleration vector $\dot{u}_a=u_{a;b}u^b$ are found as
\begin{eqnarray}
\Theta&\equiv &u^a_{\:\,;a} = \frac{1}{\surd 2\,R^2}(1+3mr^2), \\
\dot{u}_a&=&-\frac{2r}{2\sqrt{2}R^2R^2}\{(1-3mr)R^2-\Delta\}v_{a}
-\frac{a^2\sin\theta\cos\theta }{\sqrt{2}R^2R^2}
\{R\bar{m}_{a}+\bar{R}m_a\}, \nonumber
\end{eqnarray}
where $v_a={1\over\surd 2}(\ell_a-n_a)$ and $\Delta =r^2-{2mr^3}+a^2$.
Then the shear $\sigma_{ab}$  and vorticity $w_{ab}$ tensors  are respectively obtained as follows
\begin{eqnarray}
\sigma_{ab}
&=&u_{(a;b)}-\dot{u}_{(a}u_{b)}-\frac{1}{3}\Theta(g_{ab}-u_au_b) \cr &=&-\frac{1}{3\sqrt{2}R^2R^2}\Big\{3r(r^2-2mr^3+a^2)-2(2r-3mr^2)R^2\Big\}
\Big\{v_a v_b-m_{(a}\bar{m}_{b)}\Big\} \cr
&&-\frac{ia\sin\theta}{\sqrt{2}\bar{R}R^2}(2r+ia\cos\theta)v_{(a}m_{b)} -\frac{ia\sin\theta}{\sqrt{2}R\,R^2}(2r-ia\cos\theta)v_{(a}\bar{m}_{b)},\\
w_{ab}&=&u_{[a;b]}-\dot{u}_{[a}u_{b]} \cr\cr
&=&\frac{1}{\sqrt{2}R^2R^2}a^2\sin\theta\cos\theta\{R\,v_{[a}m_{b]}
+\bar{R}\,v_{[a}\bar{m}_{b]}\}\cr\cr
&&+\frac{ia\cos\theta}{\sqrt{2}R^2R^2}(\Delta+2R^2)m_{[a}m_{b]},
\end{eqnarray}
which are orthogonal to $u^a$ ({\it i.e.,} $\sigma_{ab}u^b=0$ and $w_{ab}u^b=0)$.  However, the mass $m$ directly determines the expansion $\Theta$ as well as the acceleration $\dot{u}_a$ of the solution as seen in (2.43). The existence of the vorticity tensor is due to the presence of the rotational parameter $a$ in (2.45), and confirms that the solution (2.3) is a rotating stationary solution with accelerated expansion (2.43). When $a$ sets to zero, the space-time will become the non-rotational solution (1.2) with vanishing the vorticity tensor. The existence of the acceleration vector shows that the observer of the rotating dark matter solution discussed here follows the non-geodesic path of the time-like vector ($u_{a:b}u^b\neq 0$). This provides the key feature for the accelerated expansion of the rotating dark matter. From Equations (2.43)-(2.45) imply the proof of the Theorem 2 stated in the introduction above.

\vspace*{.15in}
{\bf 3. Conclution}
\vspace*{.15in}

In this paper we derive an exact stationary (time independent) axisymmetric solution of Einstein's field equations describing rotating dark matter whose energy-momentum tensor possesses a non-perfect fluid having the negative pressure and the non-constant parameter $w(r,\theta)$ of equation of state with minus sign. However, it takes a constant value $w(r,\theta)=-1/2$ at the poles $\theta=\pi/2$ and $\theta=3\pi/2$. The most exiting property of this solution is that the metric of the solution describes both the background space-time structure and the dynamical aspects of the gravitational field in the form of the energy-momentum tensor. That is, the mass $m$ of the rotating solution plays the role for both the curvature of the space-time ({\it non-flat}) as well as the source of the energy-momentum tensor with $T_{ab}\neq 0$ ({\it non-vacuum}) measuring the energy density and the negative pressure (2.10). In the case of rotating Kerr solution, the mass plays only the role of curvature of the space-time and cannot determine the energy-momentum tensor. Hence, the Kerr solution is {\it non-flat vacuum} space-time with $T_{ab}=0$. Here the advantage of the solution (2.3) over the Kerr is that it is a {\it non-flat} and {\it non-vacuum} rotating stationary space-time metric .

The expansion, acceleration, shear and vorticity of the 4-velocity vector fields $u_a$ suggest that the observer of the rotating dark matter follows the non-geodesic path as $u_{a;b}u^b\neq 0$. We find that the energy-momentum tensor for the rotating solution violates the strong energy condition $p \geq 0$, $\rho + p \geq 0$ due to the negative pressure (2.10) of the matter field content in the space-time. This violation indicates that the gravitational field is repulsive leading to an accelerated expansion of the rotating universe. This expansion of the dark matter solution with acceleration (2.43) is consistent with the observational data [12-14].

It is emphasized the fact that the approach of the derivation of rotating dark matter solution here is merely based on the identification of the power $n=2$ of the Wang-Wu mass function without any extra assumption. This identification of the power $n=2$ in the mass function (1.1) has considered first in Paper I as mentioned above. It is to mention here that Kerr metric can be obtained by choosing $n=0$ in the Wang-Wu mass function (1.1).
In the analysis of the rotating dark matter solution, we find that the dark energy density is only attributed from the mass of the dark matter. It is the fact that without the mass of the dark matter we cannot measure the energy density and the negative pressure of the energy-momentum tensor distribution in order to obtain the dark energy equation of state $w$ with minus sign. This means that the negative pressure and the energy density of the dark fluid associated with the energy-momentum tensor are measured by the mass that produces the gravitational field in the space-time geometry of the rotating dark matter solution. The most important property of the rotating stationary dark matter solution (2.3) is that the parameter of the dark energy equation of state becomes a function $w(r,\theta)$ (2.13) with minus sign. The functional parameter $w(r,\theta)$ provides constant value $-1/2$ at the poles $\theta=\pi/2$ and $\theta=3\pi/2$, retaining the important character of a dark energy even in the case of rotating system. These poles are important points for the rotating dark matter that the event horizon and the stationary limit are coincide at these poles, where the functional parameter of the equation of state takes the constant value $w(r,\theta)=-1/2$. This situation could not be observed in the case of the non-rotating dark matter solution of Paper I. The rotating solution (2.3) suggests that dark matter and dark energy  cannot be regarded as separate components, since the {\it dark energy} with the negative pressure is produced by the mass of the {\it dark matter}, {\it i.e.} the same mass $m$ plays the important role for the existence of both.

The rotating space-time metric appears singular when $\Delta=0$, which has three roots, one is real and other two are complex. The real root $r_1$ corresponds to an event horizon. The event horizon of the dark matter is very similar to the one surrounding a black hole. Accordingly, we find area, entropy as well as surface gravity for the horizon. The existence of dark matter horizon is consistent with the cosmological horizon of de Sitter space with constant $\Lambda$ [21], which is considered to be a common candidate of dark energy with the parameter $w= - 1$ [3-11]. This parameter $w= - 1$ of the equation of state is also true for both the cosmological constant $\Lambda$ as well as the cosmological function $\Lambda(u)$ of the {\it rotating} and {\it non-rotating} de Sitter solutions [17].

From the analysis of the dark matter solutions we find the profound changes that (i) the non-rotating dark matter solution (1.2) is {\it non-asymptotic, non-vacuum, conformally flat} metric; (ii) the rotating {\it stationary} solution (2.3) is a {\it non-asymptotic, non-vacuum, Petrov type D} in the  classification of space-times, whose one of the repeated null direction  $\ell_a$ is geodesic, shear-free, expanding $[\hat{\theta}\equiv {1\over2}\ell^a_{\,\,;a}=\frac{1}{2}(\rho^*+\bar{\rho}^*)^{2}]$ as well as non-zero twist $[\hat{\omega}^2\equiv{1\over2}\ell_{[a;b]}\ell^{a;b}
=-{1\over4}(\rho^*+\bar{\rho}^*)^{2}]$ such that $C_{{abc}[d}\ell_{k]}\ell^b\ell^c=0$, where $\rho^*$ is the spin coefficient given in (2.5). It is noted that to the best of the authors knowledge, the solution describing rotating stationary dark matter presented here is not been seen discussed before. The embedded version of the rotating solution into Kerr black hole as {\it dark matter in Kerr black hole background} may be seen elsewhere in another paper. We believe that this rotating solution may provide an example of an {\it invisible rotating dark matter}, which can produce dark energy having the equation of state parameter with minus sign in an accelerated expanding space-time geometry.

\section*{Acknowledgement}

The authors, Ibohal and Ishwarchandra appreciate the Inter-University Centre for Astronomy and Astrophysics (IUCAA), Pune for hospitality during their visit in preparing the paper. The University Grants Commission (UGC), New Delhi, supports the work of Ibohal through the File No. 31-87/2005 (SR).

\end{document}